# Nanomagnonic waveguides based on reconfigurable spin-textures for spin computing

Edoardo Albisetti[1,2,*], Daniela Petti[2,*], Giacomo Sala[2], Raffaele Silvani[3,4], Silvia Tacchi[3], Simone Finizio[5], Sebastian Wintz[5], Annalisa Caló[1], Xiaorui Zheng[1], Jörg Raabe[5], Elisa Riedo[1], Riccardo Bertacco[2]

[1]Advanced Science Research Center, CUNY Graduate Center, 85 St. Nicholas Terrace, New York, New York 10031, USA.
[2]Dipartimento di Fisica, Politecnico di Milano, 20133 Milano, Italy.
[3]Istituto Officina dei Materiali del CNR (CNR-IOM), Unità di Perugia, c/o Dipartimento di Fisica e Geologia, Perugia, Italy.
[4]Dipartimento di Fisica e Geologia, Università di Perugia, Via A. Pascoli, Perugia, I-06123, Italy
[5]Paul Scherrer Institute, 5232 Villigen PSI, Switzerland

*Correspondence to: edoardo.albisetti@polimi.it; daniela.petti@polimi.it;

**Magnonics is gaining momentum as an emerging technology for information processing. The wave character and Joule heating-free propagation of spin-waves hold promises for highly efficient analog computing platforms, based on integrated magnonic circuits[1–3]. Miniaturization is a key issue but, so far, only few examples of manipulation of spin-waves in nanostructures have been demonstrated[4–6], due to the difficulty of tailoring the nanoscopic magnetic properties with conventional fabrication techniques.**

**In this Letter, we demonstrate an unprecedented degree of control in the manipulation of spin-waves at the nanoscale by using patterned reconfigurable spin-textures. By space and time-resolved scanning transmission X-ray microscopy imaging, we provide direct evidence for the channeling and steering of propagating spin-waves in arbitrarily shaped nanomagnonic waveguides based on patterned domain walls, with no need for external magnetic fields or currents. Furthermore, we demonstrate a prototypic nanomagnonic circuit based on two converging waveguides, allowing for the tunable spatial superposition and interaction of confined spin-waves modes.**

In the last years, many concepts for spin-wave based computing have been proposed[1,2,7]. Despite this, only few implementations of magnonic nanodevices have been reported[5,8]. A major challenge is the efficient channeling and steering of spin-waves in physically patterned conduits, which so far has been achieved in micron sized elements using external fields[9–13] or using arrays of nanomagnets[14]. In the route towards nanomagnonics, the use of natural nanoconduits for the propagation of spin-waves, i.e. magnetic domain walls, is highly appealing. Recently, the concept of spin-wave channeling in straight and curved domain walls has been theoretically proposed[15–17], starting from the seminal paper by Winter[18] on domain wall excitations. The first experimental evidence for spin-wave confinement at a domain wall has been provided by means of micro Brillouin Light Scattering (µ-BLS) on a straight Néel domain wall stabilized via shape anisotropy in a micron-size rectangular element[4]. However, for realizing magnonic circuits implementing computing functionalities, a much higher control of the configuration of the spin-textures is required.

In this work, we employ thermally assisted magnetic Scanning Probe Lithography (tam-SPL)[19,20] for patterning complex spin-textures in a continuous exchange biased film and controlling the propagation of spin-waves at the nanoscale. The absence of a physical patterning and the reversibility of tam-SPL allow for developing fully reconfigurable magnonic structures with engineered functionality. Through space and time-resolved scanning transmission X-ray microscopy (STXM), we provide direct evidence for the channeling and steering of localized spin-wave modes propagating along straight and curved domain wall-based waveguides, under no applied bias magnetic field.

Furthermore, we realize a structure comprising two converging nanowaveguides, allowing for a tunable interaction of the confined spin-wave modes. To our knowledge, this constitutes the first demonstration of the controlled propagation and interaction of spin-waves in a nanomagnonic circuit, paving the way to the use of engineered spin-textures as building blocks of nanomagnonic computing architectures.

In Fig. 1, we report a sketch of the experiments. Different spin-textures were patterned in an exchange bias ferromagnet/antiferromagnet bilayer (Fig. 1a), by sweeping a heated scanning probe in an external magnetic field for setting the unidirectional magnetic anisotropy strength and direction in the ferromagnetic film. This allows for the nanopatterning of engineered spin-configurations, as in the case of the curved 180° Néel domain wall of Fig. 1a, which is stabilized by patterning two magnetic domains with antiparallel remanent magnetization. Straight and curved domain walls, as well as complex spin-textures comprising two converging domain walls, were obtained by controlling the geometry of the area scanned by the tip. Spin-waves were excited by injecting an RF current in a microstrip antenna. Static and time-resolved images with magnetic contrast were acquired via STXM by measuring the transmitted X-ray intensity, so that the X-ray magnetic circular dichroism (XMCD) provides contrast to the in-plane component of the magnetization $M_x$ (Fig. 1b) (See Methods).

The configuration of the spin-wave modes is shown in Fig. 1c. Spin-waves are confined in the transverse direction of the wall by the reduced local effective field arising from the inhomogeneous magnetization profile, and propagate freely along the wall[4,15]. Such modes, called Winter magnons [15,18,21], are characterized by an elliptical precession of the spins along the wall, with the major axis lying in the film plane, associated to a propagating flexural motion of the wall profile, analogous to transverse elastic waves on a string.

Figure 2a, b shows the sample structure, consisting of an exchange bias $Co_{40}Fe_{40}B_{20}$ 20 nm / $Ir_{22}Mn_{78}$ 10 nm / Ru 2 nm multilayer[22], and the optical image of the sample. The white dashed line indicates the orientation of a patterned domain wall with respect to the antenna.

In Fig. 2c, d, e, the static STXM images of the spin-textures patterned via tam-SPL are reported, where the dark (bright) contrast corresponds to $M_x > 0$ ($< 0$). The black region at the bottom of the figure shows the boundary of the microstrip. For the straight and parabolic domain walls of Fig. 2c and Fig. 2d, respectively, the images acquired at zero external field display sharp 180° Néel walls. The corresponding micromagnetic simulations (Fig. 2f, g) show a 180° spin rotation within the sample plane, with the central spins defining the domain wall profile (in white in Fig. 2f, g) lying along the $y$-axis[19,23].

Figure 2e shows the static image of a complex spin-texture comprising two 180° Néel walls tilted by a 30° angle from each other, sharing a common apex. In this case, a static 1.5 mT magnetic field was applied in the $x$-direction in order to precisely control the distance between the two domain walls and the position of the apex (see discussion of Fig. 5). The corresponding micromagnetic simulation (Fig. 2h) shows that the two domains walls merge at the intersection, where the magnetization orientation is determined by the spin configuration of the two walls. After the intersection, a narrow "transition" region is formed, where the magnetization rotates continuously until a uniform magnetization orientation is reached within the domain.

Spin-waves were imaged stroboscopically via time-resolved STXM (see Methods). In Fig. 3, the results for straight and curved domain walls are reported. Figure 3a shows snapshots of the normalized $M_x$ contrast for the straight wall, calculated as the magnetic deviation $\Delta M_x(t)$ from the time-averaged state $\langle M_x(t) \rangle$, acquired at different times within a single period[8,24]. A Gaussian filtering was used for enhancing the contrast. The excitation frequency was 1.28 GHz and no static external magnetic field was applied. The normalized $M_x$ contrast shows spin-waves confined at the domain wall and propagating away from the antenna located at the bottom of the panels.

The spatial map of the amplitude of the spin-wave excitation is reported in Fig. 3b. The map is obtained by fitting the time-trace of each acquired pixel (raw data) with a sinusoidal function, and plotting for each pixel the amplitude of the corresponding fit. Both Fig. 3b and the horizontal profile extracted from it at $x = 1.1$ μm from the stripline (Fig. 3c) show that the mode is confined at the domain wall, with a lateral extension (FWHM) of 120 nm. In order to demonstrate the propagating character of the spin-waves, the time-traces (sinusoidal fitting) related to pixels located within the domain wall at different distances from the antenna are plotted as a function of time (Fig. 3d). The time delay observed when moving away from the antenna corresponds to a linear phase shift with distance (see Fig. 3e), clearly confirming the propagating character of the mode and allowing to estimate its wavevector.

Spin-waves propagating along a curved path, at remanence, are shown in the snapshots of Fig. 3f. The excitation frequency was 1.11 GHz. A Gaussian filtering was used for enhancing the contrast. The mode is confined at the wall and follows the wall profile, showing a lateral extension of 115 nm (see Fig. 3g, h). Both the sinusoidal fits (of raw data) as a function of time (Fig. 3i) and the phase analysis of Fig. 3l confirm the propagating nature of the mode.

Micromagnetic simulations of the confined spin-wave modes are presented in Fig. 4 (see Methods for details). Figure 4a (c) shows micromagnetic simulations for the straight (curved) wall, carried out at remanence, driven by a line source of sinusoidal magnetic field at 1.28 GHz (1.11 GHz), as in the experimental data of Fig. 3. The $M_x$ and $M_z$ components are reported in the left and right panels, respectively, while the magnetic deviation $\Delta M_x(t)$ from the time averaged value $\langle M_x(t) \rangle$, which corresponds to the STXM signal reported in Fig. 3, is shown in the central panel. The flexural motion of the wall, associated with the spin precession within the wall, can be observed for both straight and curved geometries. In good agreement with the experimental results, these findings indicate that wall-bound Winter-like modes, propagating along the wall, can be excited in both the straight and curved geometries.

Figure 4b reports the dispersion relation simulated for spin-waves confined at a straight 180° Néel domain wall, together with the experimental one. The simulations were performed with the same geometry as in Fig. 4a, but using a sinc-shaped field pulse as excitation (see Methods). The experimental values of the wavevector were extracted from the linear fitting of the phase shift vs. distance curves obtained from experiments performed at different excitation frequencies (see discussion for Fig. 3).

In good agreement with the experimental results, the simulations confirm the propagating character of the excitations, showing a positive dispersion and the presence of a small bandgap below 0.3 GHz, which can be ascribed to the residual effective field within the domain wall due to exchange bias and uniaxial anisotropies[15]. The simulated (experimental) spin-wave group velocity $v_g = \partial\omega/\partial k$, extracted at $k \to 0$, is $v_{g\text{Sim}} = 2.30 \pm 0.34$ km/s ($v_{g\text{Exp}} = 2.77 \pm 1.4$ km/s). Noteworthy, the demonstration of the propagating character and of the positive dispersion confirms that such guided modes can be used for transporting information within integrated nanomagnonic circuits.

Waveguides for efficiently controlling and manipulating confined spin-wave modes constitute fundamental building blocks for the realization of nanomagnonic devices. In the following, we demonstrate a nanomagnonic circuit allowing for the tunable spatial superposition and interaction of guided spin-wave modes propagating in two converging waveguides.

Figure 5a shows the STXM images of a spin-texture comprising two domain walls (see also Fig. 2e). By applying a small static magnetic field in the $-x$ direction, ranging from 2 mT to 1.68 mT, the distance between the two domain walls (dashed white lines) can be controlled. In the top panel, the two domain walls are spatially separate. By decreasing the field, the two walls are brought closer (central panel) and finally converge at the common apex (lower panel).

Figure 5b shows STXM snapshots from the three different configurations for a 1.28 GHz excitation frequency. The images were smoothed with a gaussian filter for increasing contrast. The normalized $M_x$ contrast shows, in all three cases, two guided spin-wave modes propagating from the antenna with different relative phases. These two modes, which are spatially separated close to the antenna, approach as the domain walls converge, and partially overlap for low applied fields. In order to better visualize the progressive overlapping of the two modes, each pixel of the data of Fig. 5b was fitted with a sinusoidal function (see the discussion of Fig. 3). Figure 5c shows the amplitude of the sinusoidal fit along the horizontal profiles of Fig. 5a (green dashed lines), in the three configurations.

The two peaks, with FWHM of around 200 nm, correspond to the two guided modes. For an applied magnetic field of 2 mT, the two modes are separated by 810 nm and do not overlap. By decreasing the field down to 1.68 mT, the two modes are brought closer and to partially overlap, with a peak-to-peak distance of 340 nm.

In the top panels of Fig. 5d the sinusoidal fits along the horizontal profiles of Fig. 5a (green dashed lines) are plotted as a function of time for the different applied fields. In the bottom panels, single sinusoidal profiles are extracted from the positions marked by the color-coded stars in the top panels. Blue and yellow curves show the magnetization dynamics in correspondence of the maximum amplitude of the two guided modes. As expected, their phase difference depends on the applied field, because of the modulation of the waveguide geometry and spin-configuration.

The red curves show the dynamics in the region where the two modes overlap. For 2.00 mT (left panel), the two modes are spatially separate, therefore at $y = 0$ no excitation is measured (red dashed line). For lower fields (central and right panels), we observe a sizeable modulation of the excitation amplitude and phase in the overlap region, which arises from the tuning of the spatial superposition of the two guided modes. This is a compelling demonstration that nanoscale designed and actively tunable spin-textures can enable flexible nanomagnonic computing devices with reconfigurable functionality.

In conclusion, we demonstrated the realization of spin-wave nanowaveguides based on patterned reconfigurable spin-textures. We directly imaged and studied the channeling and steering of spin-waves propagating along straight and curved paths, without the need for external applied fields. The experimental demonstration of the spin-wave guidance along nanopatterned arbitrary shaped channels opens up a plethora of new possibilities in the field of spin computing. In this framework, we realized a prototypical nanomagnonic circuit based on two converging waveguides, where we control the spatial superposition and interaction of spin-waves by reconfiguring the spin-texture via external stimuli. Our results pave the way to the development of energy-efficient computational architectures based on the manipulation of propagating spin-waves in reconfigurable nanomagnonic circuits.

**Acknowledgements**

The research leading to these results has received funding from the European Union's Horizon 2020 research and innovation programme under grant agreements no. 705326, project SWING, and no. 730872, project CALIPSOplus. This work was partially performed at Polifab, the micro- and nano-technology center of the Politecnico di Milano. Part of this work was performed at the PolLux (X07DA) endstation of the Swiss Light Source, Paul Scherrer Institut, Villigen, Switzerland. The PolLux endstation was financed by the German Minister für Bildung und Forschung (BMBF) through contracts 05KS4WE1/6 and 05KS7WE1.

**Authors contribution**

E.A. and D.P. conceived and designed the experiments. E.R. and R.B. supervised the research work. E.A. performed the tam-SPL patterning and data analysis. D.P. and G.S. fabricated the samples. E.A., D.P., G.S., S.F., S.W. and J.R. performed the STXM measurements. E.A., R.S. and S.T. performed the simulations. E.A., D.P., S.T. and R.B. wrote the manuscript. All the authors contributed to discussions regarding the research.

**Methods**

**Sample fabrication**. $Co_{40}Fe_{40}B_{20}$ 20 nm / $Ir_{22}Mn_{78}$ 10 nm / Ru 2 nm stacks were deposited on 200 nm thick $Si_3N_4$ membranes by DC magnetron sputtering using an AJA Orion8 system with a base pressure below $1\times10^{-8}$ Torr. During the deposition, a 30 mT magnetic field was applied in the sample plane for setting the magnetocrystalline uniaxial anisotropy direction in the CoFeB layer and the exchange bias direction in the as-grown sample. Then, the samples underwent an annealing in vacuum at 250 °C for 5 minutes, in a 400 mT magnetic field oriented in the same direction as the field applied during the growth. The resulting exchange bias field was 2.5 mT.

Thermally assisted magnetic Scanning Probe Lithography (tam-SPL) was performed via NanoFrazor Explore (SwissLitho AG). Spin-textures were patterned by sweeping in a raster-scan fashion the scanning probe, heated above the blocking temperature of the exchange bias system $T_B \approx 300°$ C, in presence of an external magnetic field. Two rotatable permanent magnets were employed for generating a uniform external magnetic field applied in the sample plane during patterning.

2 μm × 30 μm microstrip antennas were then fabricated via optical lithography using a Heidelberg MLA100 Maskless Aligner and lift-off, after depositing a 50 nm thick $SiO_2$ insulating layer via magnetron sputtering. A Cr 5 nm / Cu 200 nm bilayer was deposited by means of thermal evaporation.

**Scanning trasmission X-ray microscopy**. The time-dependent magnetic configuration of the samples was investigated with time-resolved scanning transmission X-ray microscopy at the PolLux (X07DA) endstation of the Swiss Light Source[25]. In this technique, monochromatic X-rays, tuned to the Co $L_3$ absorption edge (photon energy of about 781 eV), are focused using an Au Fresnel zone plate with an outermost zone width of 25 nm onto a spot on the sample, and the transmitted photons are recorded using an avalanche photodiode as detector. To form an image, the sample is scanned using a piezoelectric stage, and the transmitted X-ray intensity is recorded for each pixel in the image. The typical images we employed for the investigation of the spin-wave propagation in our samples were acquired with a point resolution between 40 nm and 75 nm.

Magnetic contrast in the images is achieved through the X-ray magnetic circular dichroism (XMCD) effect, by illuminating the sample with circularly polarized X-rays. As the XMCD effect probes the component of the magnetization parallel to the wave vector of the circularly polarized X-rays, the samples were mounted to achieve a 30° orientation of the surface with respect to the X-ray beam, allowing us to probe the in-plane component of the magnetization.

The time-resolved images were acquired in a pump-probe scheme, using an RF magnetic field generated by injecting an RF current in a microstrip antenna as pumping signal and the X-ray flashes generated by the synchrotron light source as probing signal. The pumping signal was synchronized to the 500 MHz master clock of the synchrotron light source (i.e. to the X-ray flashes generated by the light source) through a field programmable gate array (FPGA) setup. Due to the specific requirements of the FPGA-based pump-probe setup installed at the PolLux endstation, RF frequencies of $f_{\text{exc}} = 500 \times M/N$ [MHz], being $N$ a prime number and $M$ a positive integer, were accessible. For the measurements presented in this work, $N$ was typically selected to be equal to 23, giving a phase resolution of about 15° in the time-resolved images. Depending on the RF frequency, the temporal resolution of the time-resolved images is given by $2/M$ [ns], with its lower limit given by the width of the X-ray pulses generated by the light source (i.e. about 70 ps FWHM).

**Micromagnetic simulations.** Micromagnetic simulations of the magnetization dynamics were carried out by solving the Landau-Lifshitz-Gilbert equation of motion, using the open-source, GPU-accelerated software MuMax3. The total simulated volume had dimensions of 20480×2560×20 nm³ and of 10240×5120×20 nm³ for the straight and the curved wall, respectively, and was discretized into cells having dimensions of 5×5×20 nm³. Periodic boundary conditions in the $x$-direction were used to reproduce an infinite domain wall. The following parameters for the CoFeB were used: saturation magnetization $M_s = 800$ kA/m, in-plane uniaxial anisotropy constant $K_u = 10^3$ J/m³ with the easy direction parallel to the $x$-axis (see Fig. 4 in the main text) and exchange constant $A_{ex} = 2 \cdot 10^{-11}$ J/m. The Gilbert damping parameter was set to $\alpha = 0.02$. The exchange bias field was modeled as an external magnetic field of 2.5 mT, applied along the $x$-axis in opposite direction inside and outside the patterned area. In order to simulate the transition between two domains with opposite exchange bias, a 250 nm wide transition region with zero exchange bias field was placed in correspondence of the domain wall.

To simulate the spatial profile of the spin-wave modes, for both the straight and the curved wall the magnetization dynamics was excited applying a time-varying sinusoidal magnetic field to one line of cells at the center of the rectangular region and parallel to the $y$-axis. The field amplitude was 30 mT. In the simulation of the dispersion relation of the straight wall, in order to excite spin-waves, we used a sinc-shaped field pulse $b(t) = b_0 \frac{\sin(2\pi f_0 (t-t_0))}{2\pi f_0 (t-t_0)}$ directed along the $x$-axis, with amplitude $b_0 = 1$ T and frequency $f_0 = 5$ GHz. The dispersion relation was calculated by performing a Fourier-transform of the $x$-component of the magnetization both in space and time in the whole simulated area.

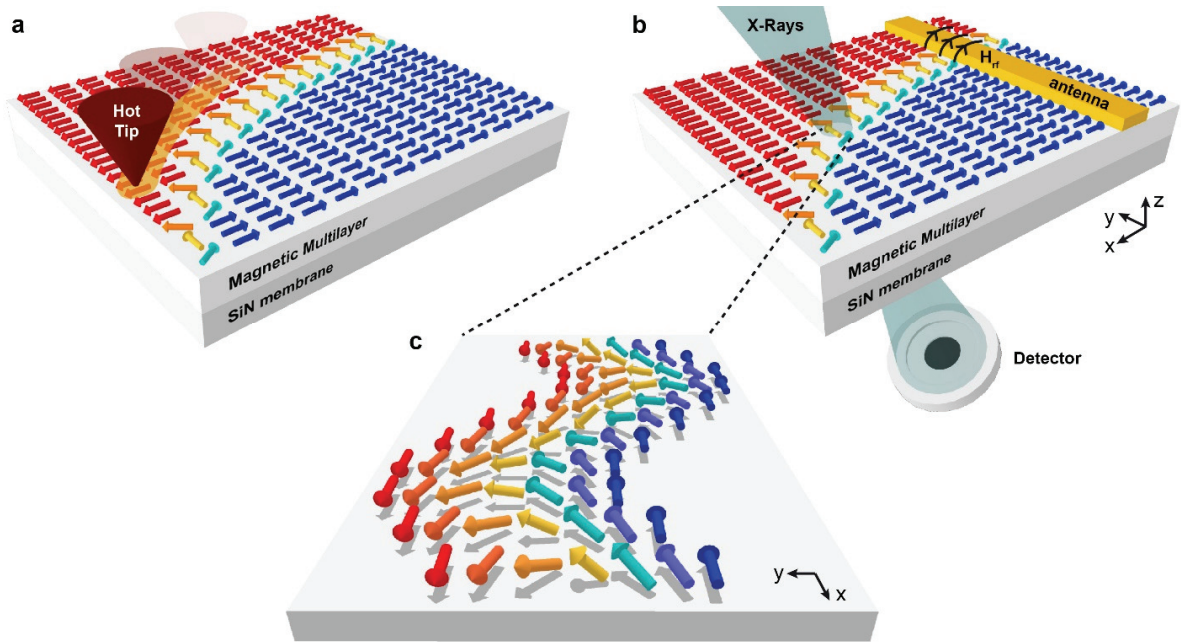

**Figure 1. Patterning of nanoscale spin-textures and study of the confined spin-wave modes. a**, Nanoscale spin-textures, such as magnetic domain walls with tailored spin configuration, are patterned via thermally assisted magnetic Scanning Probe Lithography (tam-SPL) in a continuous exchange biased ferromagnetic layer. **b**, The static characterization of the patterned spin-textures and the study of the localized spin-wave modes are performed via scanning transmission X-ray microscopy (STXM). A microstrip antenna is employed for the spin-wave excitation. **c**, Schematic representation of a spin-wave mode confined at a 180° Néel domain wall, propagating freely along the wall.

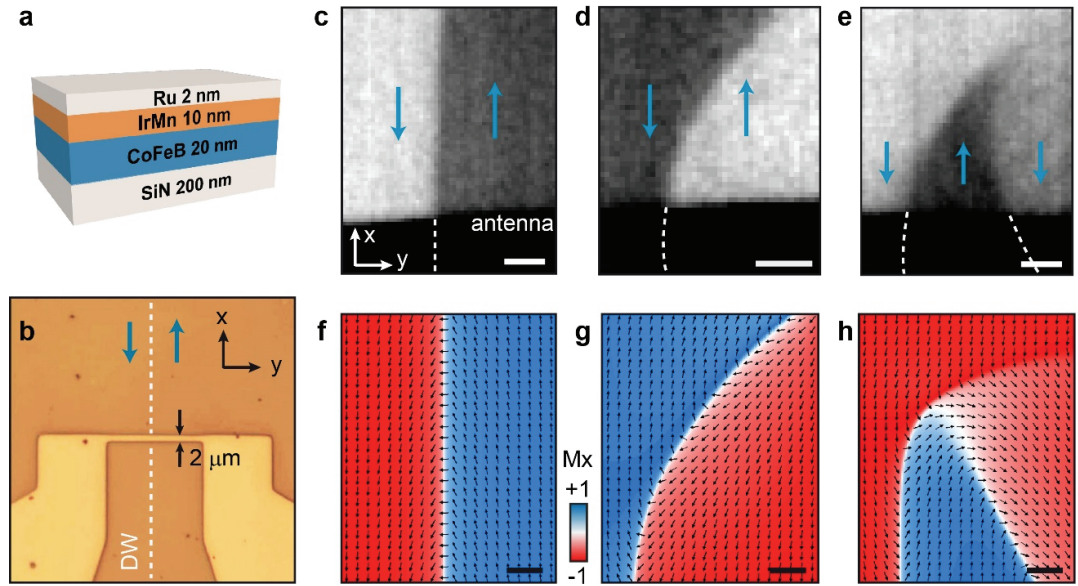

**Figure 2. Sample structure, static characterization and micromagnetic simulations of the patterned spin-textures. a**, The tam-SPL patterning and the STXM study were performed on a CoFeB 20 nm / IrMn 10 nm continuous exchange bias bilayer. **b**, Optical image of the sample, showing a 2 μm wide microstrip antenna for spin-wave excitation. The white dashed line indicates the position of a straight domain wall patterned via tam-SPL, with respect to the antenna. Blue arrows indicate the direction of the CoFeB magnetization within the domains. **c**, **d**, **e**, Static STXM images of a straight 180° Néel domain wall, a parabola-shaped 180° Néel wall, and a complex spin-texture comprising two converging 180° Néel walls sharing a common apex, respectively. The black (white) contrast indicates $M_x > 0 \, (< 0)$, confirming the stabilization of sharp Néel walls. The arrows indicate the direction of the magnetization within the domains. The dashed lines indicate the continuation of the domain walls under the antenna. No external magnetic field was applied for panels **c**, **d**. In panel **e**, a 1.5 mT field was applied along $+x$ for controlling the distance between the two domain walls and the position of the apex. **f**, **g**, **h**, Micromagnetic simulations corresponding to the upper panels. The black arrows indicate the local spin-configuration. The blue (red) color indicates $M_x > 0 \, (< 0)$. Scale bars 500 nm.

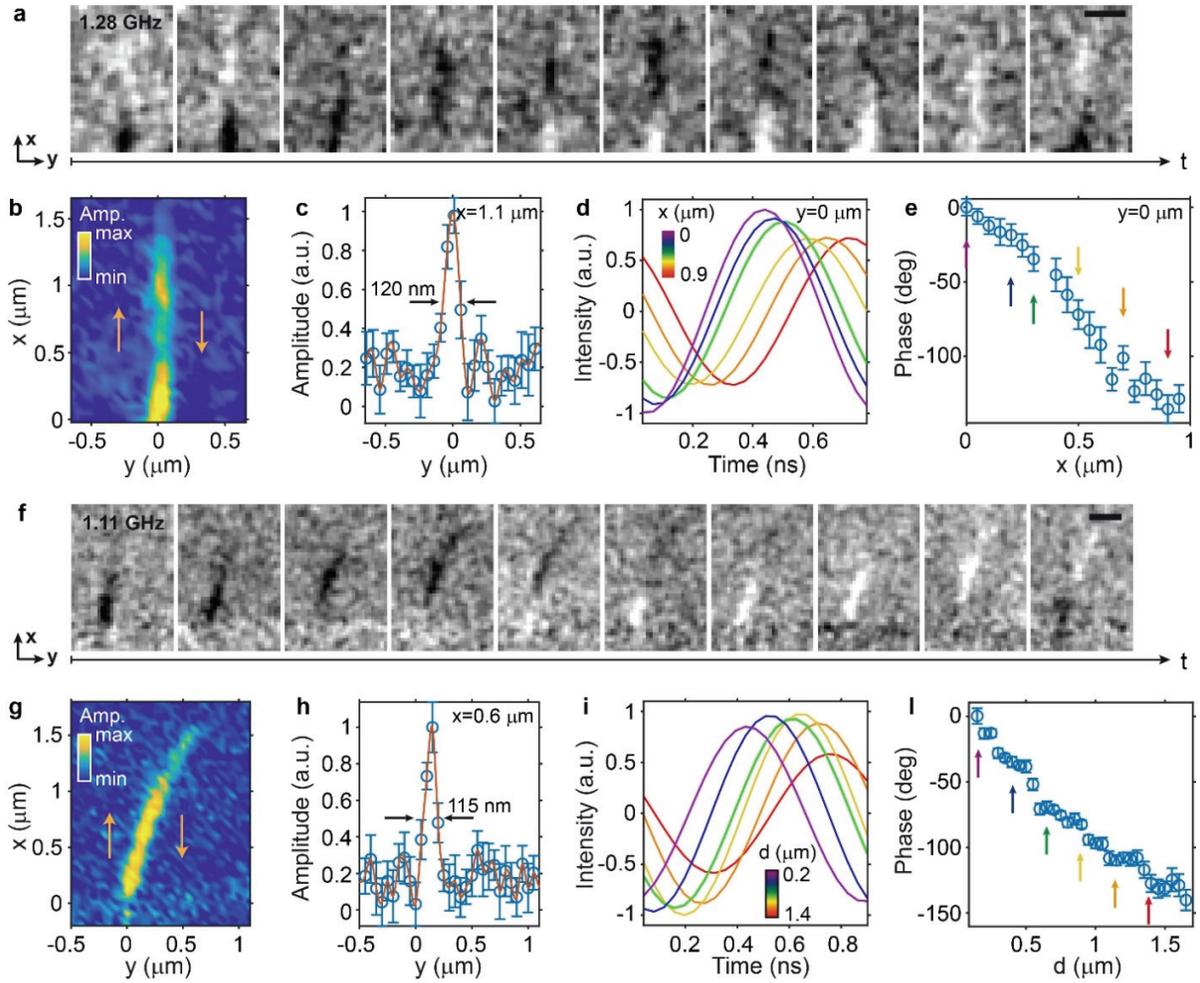

**Figure 3. Localized spin-waves propagating along straight and curved domain walls. a, f,** Normalized time-resolved STXM images showing the propagation of localized spin-waves along a straight 180° Néel wall (**a**) and a parabola-shaped 180° Néel wall (**f**). The frequency of the excitation current was 1.28 GHz (1.11 GHz) for the straight (curved) wall. **b, g,** Spatial maps of the spin-wave excitation amplitude, showing localization at the domain wall for the straight (**b**) and curved (**g**) wall. The arrows indicate the direction of the magnetization within the domains. The $x = 0, y = 0$ position is set in correspondence of the wall, right outside the antenna. **c, h,** Horizontal profiles extracted from panel **b** at $x = 1.1$ μm from the antenna (**c**), and from panel **g** at $x = 0.6$ μm from the antenna (**h**). The FWHM of the excitation peaks in correspondence of the domain wall is 120 nm (**c**) and 115 nm (**h**). **d, i,** STXM time traces (sinusoidal fitting) showing the excitation at different distances from the antenna, color-coded, along the domain wall for the straight (**d**) and curved (**i**) wall. **e, l,** Phase shift as a function of distance from the antenna along the straight (**e**) and curved (**l**) domain wall confirming the propagating character of the spin-waves. Color-coded arrows indicate the corresponding time traces of panel **d** and **i**, respectively. Scale bars: 500 nm.

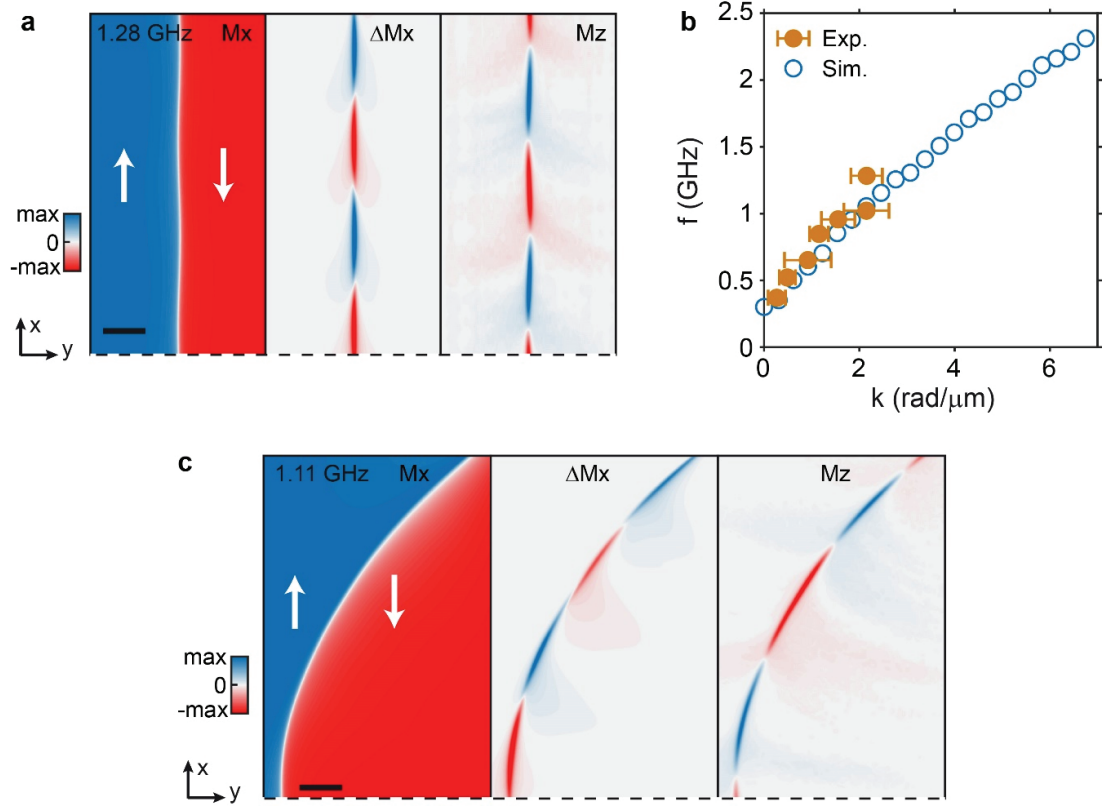

**Figure 4. Spin-wave modes, simulated and experimental dispersion. a**, **c**, Micromagnetic simulations of the spin-waves mode in presence of a straight Néel 180° domain wall (**a**) and in presence of a curved Néel 180° wall (**c**). The excitation (frequency indicated in each panel) is generated by a horizontal line antenna placed below the dashed line. From the left to the right, the $M_x$, the $\Delta M_x$ and the $M_z$ are shown. The mode is characterized by the flexural motion of the domain wall in the plane of the film and by the precession of the spins along the domain wall. The arrows indicate the direction of the magnetization within the domains. Scale bars: 500 nm. **b**, Dispersion of the spin-waves confined at the straight domain wall. Empty blue (filled orange) circles represent the simulated (experimental) dispersion.

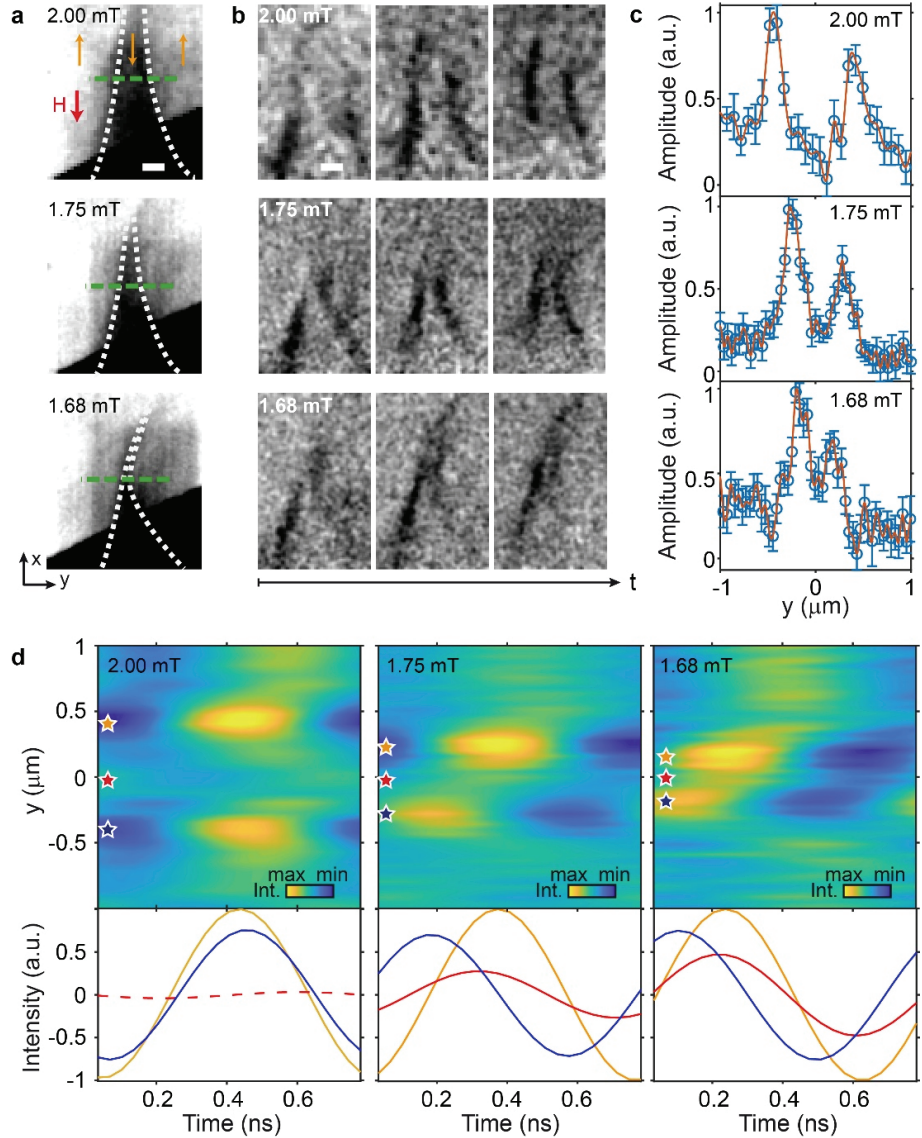

**Figure 5. Tunable spatial superposition of two modes via external fields. a**, STXM images of a spin-texture consisting of two 180° Néel domain walls sharing a common apex, for different external magnetic fields applied in the $-x$ direction (red arrow). The distance between the two domain walls, marked by dashed white lines, is controlled by the external field. Orange arrows indicate the direction of the magnetization within the domains. **b**, Normalized time-resolved STXM images showing spin-waves excited at 1.28 GHz propagating along the two domain walls, for different external fields, indicated at the top of each panel. **c**, Spin-wave excitation amplitude as a function of the distance along the horizontal profiles indicated by the green dashed lines in panel **a**, for different external fields. **d,** In the top panels, STXM time traces (sinusoidal fitting) as a function of $y$, along the horizontal profiles marked by the green dashed lines in panel **a**, for different external fields. In the bottom panels, time-traces extracted in correspondence of the color-coded stars in the panel above.

For 2 mT the modes are spatially separate. By reducing the external field, a tunable spatial superposition of the two modes is achieved as shown by the increasing excitation amplitude at $y = 0$ μm for 1.75 mT and 1.68 mT. Scale bars: 500 nm.